\documentclass[9pt,conference]{IEEEtran}
\usepackage{graphicx}
\usepackage{float}
\usepackage{amsfonts}
\usepackage{amsmath}
\usepackage{amssymb}
\usepackage{cite}
\usepackage{bm}
\usepackage{mathrsfs}
\usepackage{epsfig}
\usepackage{algorithm}
\usepackage{algorithmic}
\usepackage{amsthm}
\usepackage{color}
\usepackage{graphicx}
\usepackage{caption}
\usepackage{subcaption}

\begin{document}

\title{Robust Detection of Random Events with Spatially Correlated Data in Wireless Sensor Networks via Distributed  Compressive  Sensing }

\author{\authorblockA{Thakshila Wimalajeewa  \emph{Member IEEE}, and Pramod K.
Varshney, \emph{Fellow IEEE}}
}

\maketitle\thispagestyle{empty}

\begin{abstract}
In this paper, we exploit the theory of compressive sensing to perform detection of a random source in a dense sensor network. When the  sensors are densely deployed, observations at adjacent sensors are highly correlated while those corresponding to  distant sensors are  less correlated. Thus, the covariance matrix of the concatenated observation vector of all the sensors at any given time  can be sparse where the sparse structure depends on the network topology and the correlation model. Exploiting the sparsity  structure of the covariance matrix, we develop  a robust nonparametric detector to detect the presence of the random event using  a compressed version of the data collected  at the distributed  nodes. We employ  the multiple access channel (MAC)  model with distributed random projections  for sensors to transmit observations so that  a compressed version of the observations  is available  at  the fusion center. Detection is performed by constructing a  decision statistic based on the covariance information of uncompressed data which is estimated using compressed data.  The proposed approach does not require any knowledge of the noise  parameter to set the threshold, and is also   robust when the  distributed random  projection matrices become sparse.
\end{abstract}

\footnotetext[1]{This work was supported in part by ARO grant no. W911NF-14-1-0339.  The authors are with the Dept. EECS, Syracuse University, Syracuse, NY. Email: \{twwewelw,varshney\}@syr.edu }

\section{Introduction}
Over the last two decades, wireless sensor network (WSN) technology has  gained increasing attention by both research community and actual users \cite{Akyildiz_CN2002,Puccinelli_MCAS2005,Yick_CN2008,Mainetti_SoftCOM2011,
Stankovic_CSN2011,Othman_EP2012,Rawat_JoS2014,Rashid_JNCA2016}. Sensor networks are inherently resource constrained and they starve for energy and communication efficient  protocols \cite{Akyildiz_CN2002}. There is  abundant literature related  to energy-saving in
WSNs as numerous methods have been proposed for energy efficient protocols in the
last several  years. However, there is still much ongoing research
on how to optimize power and communication bandwidth  in resource constrained sensor
networks since none of the existing standalone protocol is universally applicable.

Recent advances in compressive sensing (CS) have  led to novel approaches  to design energy efficient WSNs. 
Sparsity is a common characteristic that can be observed in WSN applications  in various  forms. For example, in many applications, the time samples collected at a given  node can be represented in a sparse manner in a given basis \cite{Duarte_TIT13}. When considering  multiple measurement vectors (MMVs) collected at distributed nodes, different sparsity patterns with certain structures can be observed \cite{Duarte_TIT13}. Joint processing of such MMVs using CS techniques  by exploiting  temporal sparsity along with different joint structures leads to energy efficient  signal processing as desired by WSNs.  Spatial sparsity of observations collected at distributed nodes is another form of sparsity. For example,  since not all the  sensors  gather informative observations at any given time, to make a compressed version of the observations available at the fusion center, random projections can be employed  \cite{haupt_SPM2008}.
Spatial sparsity can  also be leveraged  by construction such as in source localization and sparse event  detection \cite{meng_CISS09,Feng_Globecom2009,Zhang_Infocom2011}. In addition to complete signal reconstruction as is commonly done in the CS literature, CS has been exploited for detection problems exploiting temporal, or  spatial spatial sparsity \cite{duarte_ICASSP06,haupt_ICASSP07,Gang_globalsip14,Rao_icassp2012,Cao_Info2014,Wimalajeewa_tsipn16} or without exploiting any sparsity prior of signals \cite{davenport_JSTSP10,Wimalajeewa_asilomar10,Bhavya_cscps14,Bhavya_asilomar14}.

In contrast to these existing works, in this paper, our goal is to exploit the sparsity or structural  properties of the covariance matrix of spatially correlated data (but not sparsity of observations itself) to solve a random event detection problem. In particular, a decision statistic is computed using the covariance information of data collected at multiple sensors.  In a typical WSN, the densely deployed sensor observations can be highly correlated.  In \cite{Vuran_Else2004,Berger_Stat2001}, several spatial correlation models have been discussed. With most of these models, the correlation among nodes that are located  far from each other is negligible. Thus, the covariance matrix of the concatenated data vector can have a sparse or some known structure which is determined by the spatial correlation model and the network topology. If only a compressed version of the concatenated data vector is received at the fusion center, the covariance matrix can be computed based on compressed data as considered in compressive covariance sensing \cite{Romero_SPM16}. To have a compressed version of spatially correlated data at the fusion center, we employ the multiple access channel (MAC) model with distributed random projections  \cite{haupt_SPM2008,Bajwa_IT2007}. Using the sample estimate of the covariance matrix of compressed data with limited samples, we compute a decision statistic in terms of  the covariance  matrix of uncompressed data. The proposed approach does not require any knowledge of  the noise  parameters for threshold setting as needed by likelihood ratio (LR) based and/or   energy detectors. Further, the proposed approach is shown to be robust to the selection of the distributed projection matrices (i.e., dense vs sparse matrices).

This work is motivated by our  recent work in \cite{Wimalajeewa_TSP17}, in which a similar decision statistic  was computed to perform detection with multi-modal (non-Gaussian in general) dependent data in the compressed domain. However, the application scenario and the problem formulation in this work are  different from that in \cite{Wimalajeewa_TSP17} mainly with respect to the compression model used at each sensor  and the communication architecture between the sensors and the fusion center.

\section{Detection with Spatially Correlated Data in WSNs}\label{sec_formulation}
Let there be $L$ sensor nodes in a network deployed to solve a binary hypothesis testing  problem where the  two hypotheses are denoted by $\mathcal H_1$ (signal present) and $\mathcal H_0$ (signal absent). Consider the detection of a random signal, denoted by $S$, emitted by a point source. The $n$-th measurement at the $j$-th node is denoted by $x_{nj}$ for $j=1,\cdots,L$ and $n=1,\cdots,T$. Under the two hypotheses, $x_{nj}$  is given by.
\begin{eqnarray}
\mathcal H_1&:& x_{nj}  = s_{nj}+ v_{nj}\nonumber\\
\mathcal H_0&:& x_{nj}  =  v_{nj}\label{obs_0}
\end{eqnarray}
for $j=1,\cdots,L$ and $n=1,\cdots, T$,
where $s_{nj}$ is the  realization  of $S$ at the $j$-th node at time $n$, $ v_{nj}\sim\mathcal N(0,\sigma_v^2)$ is the noise which is assumed to be Gaussian and iid    over $j$ and $n$.
We further define $\mathbf x[n]=[x_{n1}, \cdots, x_{nL}]^T$ to be the observation vector over all the nodes at time $n$. Similarly, we use the notations $\mathbf s[n]$ and $\mathbf v[n]$, respectively, to denote the signal and noise vectors at time $n$. The mean and the variance of $S$ are denoted by $\mu_S$ and $\sigma_S^2$, respectively.  Without loss of generality,   we assume that $\mu_S=0$.

In a dense sensor network where  the nodes are located very close to each other,  the elements of $\mathbf s[n]$ can be correlated at any given time when  all the nodes observe the same random phenomenon.  Let the covariance matrix of $\mathbf s[n]$ be denoted by $\boldsymbol\Sigma_s$ with the $(i,j)$-th element, $\boldsymbol\Sigma_s[i,j] = \rho_{ij} \sigma_{S}^2$ for $i\neq j$.
We define $\rho_{ij}$ to be the correlation coefficient between $s_{ni}$ and $s_{nj}$ which is given by
\begin{eqnarray}
\rho_{ij} = \frac{\mathrm{cov}(s_{ni}, s_{nj})}{\sigma_{S}^2}.
\end{eqnarray}
In \cite{Vuran_Else2004}, several spatial correlation models were discussed in which  $\rho_{ij}$ is expressed as $\rho_{ij} = G_{\vartheta}(r_{ij})$ where $r_{ij}$ denotes the distance between the $i$-th node  and the $j$-th  node, and $G_{\vartheta}(\cdot)$ defines the correlation model (e.g., spherical, power exponential, etc..).   If $S$ is assumed to be Gaussian and $\boldsymbol\Sigma_s$ and $\sigma_v^2$ are known, the LR  test  can be employed  to solve the detection problem \eqref{obs_0} assuming that $\mathbf x[n]$ for $n=1,\cdots, T$ is available at a central fusion center. However, when these parameters are unknown and/or $S$ is not Gaussian, performing LR based detection is challenging. In such scenarios, one of the commonly used nonparametric detectors is the energy detector. While the energy detector shows good performance when $S$ is Gaussian, its susceptibility to  the exact knowledge of the noise power makes the energy detector not very attractive in many practical settings. Further, making  $\mathbf x[n]$ available at the fusion center may require considerable communication overhead which can be undesirable in   resource constrained sensor networks.

To address these issues, we exploit  CS theory  to make  a compressed version of $\mathbf x[n]$ available at the fusion center and propose a robust  nonparametric detector based on covariance information of the uncompressed observations. When  the random event is present, the covariance matrix of $\mathbf x[n]$ is non-diagonal  while it is diagonal in the presence of only noise. Thus, a decision statistic based on the covariance matrix of $\mathbf x[n]$ can be used to perform detection.  On the other hand, based on most of the spatial correlation models discussed in \cite{Vuran_Else2004}, the observations at nearby sensors are strongly correlated while the correlation reduces as the distance between nodes increases. Thus,   $\boldsymbol\Sigma_s$  can be assumed to have  a sparse structure. If a compressed version of $\mathbf x[n]$ is available at the fusion enter, the concepts of CS can be utilized to construct a decision statistic based on  $\boldsymbol\Sigma_s$ without having access to  the raw observations $\mathbf x[n]$.

\section{Nonparametric Compressed Detection of a Random Event  via MAC}
To obtain a compressed version of $\mathbf x[n]$ at the fusion center,  we employ the MAC architecture as proposed in \cite{haupt_SPM2008,Bajwa_IT2007}. In  the MAC model, the $j$-th node multiplies its observation at time $n$ by a scalar quantity denoted by $\mathbf A[i,j]$ and transmits it  coherently so that the fusion center receives,
\begin{eqnarray}
y_{ni} = \sum_{j=1}^L \mathbf A[i,j] \mathbf x_{nj} + w_{ni}
\end{eqnarray}
with the $i$-th transmission
where $w_{ni}\sim \mathcal N(0,\sigma_w^2)$ is the noise at the fusion center which is assumed to be Gaussian and  iid. The observed signal vector at the fusion center at time $n$ after $M$ transmissions can be expressed  as
$
\mathbf y[n] = \mathbf A \mathbf x[n] + \mathbf w[n]
$
where $\mathbf A\in \mathbb R^{M\times N}$, and $\mathbf w[n] \sim \mathcal N(\mathbf 0,\sigma_w^2\mathbf I)$ with $\mathbf I$ denoting the identity matrix. With this model, the detection problem reduces to,
\begin{eqnarray}
\mathcal H_1: ~ \mathbf y[n] &=& \mathbf A \mathbf s[n] + \tilde{\mathbf w}[n]\nonumber\\
\mathcal H_0: ~ \mathbf y[n] &=&  \tilde{\mathbf w}[n]\label{obs_MAC}
\end{eqnarray}
where $\tilde{\mathbf w}[n] = \mathbf A \mathbf v[n] + \mathbf w[n]$. In the rest of the paper, we assume that the elements of $\mathbf A$ are zero mean random and satisfy $\mathbf A \mathbf A^T = \mathbf I$ (we discuss the robustness of the proposed method when this condition is relaxed in  Section  \ref{sec_numerical}).  Then, we have $\tilde{\mathbf w}[n]\sim \mathcal N(\mathbf 0, \sigma_{\tilde w}^2 \mathbf I)$ where $\sigma_{\tilde w}^2 = \sigma_v^2 + \sigma_w^2$. Let $\boldsymbol\Sigma_y =\mathbb E\{\mathbf y[n] \mathbf y[n]^T\}$ denote the covariance matrix of $\mathbf y[n]$ which is given by $\boldsymbol\Sigma_y =  \mathbf A \tilde{\boldsymbol\Sigma}_x \mathbf A^T$ where
\begin{eqnarray}
\tilde{\boldsymbol\Sigma}_x=\left\{
\begin{array}{ccc}
\boldsymbol{\Sigma}_s + \sigma_{\tilde w}^2 \mathbf I \triangleq \tilde{\boldsymbol\Sigma}_s~& \mathrm{under} ~ \mathcal H_1\\
\sigma_{\tilde w}^2 \mathbf I  ~& \mathrm{under} ~ \mathcal H_0
\end{array}\right..\label{cov_y}
\end{eqnarray}
 It is noted that $\tilde{\boldsymbol\Sigma}_x$ is the covariance matrix of  $\mathbf x[n]$  if  $\mathbf x[n]$ was available at the fusion center in the presence of noise with mean zero and the covariance matrix   $\sigma_w^2\mathbf I$.

The goal is to decide as to  which hypothesis is  true based on \eqref{obs_MAC} when the signal and noise statistics are completely unknown at the fusion center.  From \eqref{cov_y}, it is seen that $\tilde{\boldsymbol\Sigma}_x$  has different structures under the two hypotheses which can be used to construct a decision statistic.
Here we consider the following  decision statistics based on $\tilde{\boldsymbol\Sigma}_x$ \cite{Wimalajeewa_TSP17,Zeng_C2007,Zeng_VT09}:
\begin{eqnarray}
\Lambda_C = \frac{\underset{i,j}{\sum}|\tilde{\boldsymbol\Sigma}_x [i,j] |}{\underset{i}{\sum}|\tilde{\boldsymbol\Sigma}_x[i,i] |} \label{Lamda_C}
\end{eqnarray}
where $|\cdot|$ denotes the absolute value.
Note that $\boldsymbol\Sigma_y$ is a compressed version of $\tilde{\boldsymbol\Sigma}_x$ where $\tilde{\boldsymbol\Sigma}_x$ has a sparse structure under $\mathcal H_1$ with different correlation models as discussed in \cite{Vuran_Else2004}. This motivates us to exploit the concepts of compressive covariance sensing \cite{Romero_SPM16} to efficiently compute $\Lambda_C$ based on   $\boldsymbol\Sigma_y$. In this paper, we replace $\boldsymbol\Sigma_y$ by its sample estimate,  $\tilde{\boldsymbol\Sigma}_y$, which  is given by, $\tilde{\boldsymbol\Sigma}_y = \frac{1}{T} \sum_{n=1}^T \mathbf y[n]  \mathbf y[n]^T$.
\subsection{Computation of $\Lambda_C$ }
The specific procedure to estimate $\tilde{\boldsymbol\Sigma}_x$  from $\tilde{\boldsymbol\Sigma}_y$ depends on the structure of $\tilde{\boldsymbol\Sigma}_x$ which depends on the sensor network configuration  and the correlation model. Here, we consider a specific architecture  for the sensor network.

\subsubsection{Equally spaced 1D sensor network}
When the sensors in a 1-D  network are equally spaced with the node index order $[1,\cdots, L]$, with the correlations models considered in \cite{Vuran_Else2004},  ${\boldsymbol\Sigma}_s$ (and thus $\tilde{\boldsymbol\Sigma}_s$) can be assumed to have  a Toeplitz structure. Let $\mathbf d = [d_1, \cdots, d_L]$ denote  the first row of $\tilde{\boldsymbol\Sigma}_s$ which is given by $d_1=\sigma_S^2 + \sigma_{\tilde w}^2$ and $d_k=\rho_{k-1} \sigma_S^2$ for some $-1<\rho_k<1$ for $k=2,\cdots,L$. It is noted that $\tilde{\boldsymbol\Sigma}_s$ is determined by $\mathbf d$. With this  structure, estimation of $\tilde{\boldsymbol\Sigma}_x$ reduces to the estimation of $L$ unknown parameters  in general. The accuracy of the estimates depends on $M$ and $T$. Since it is desired to keep $M$ and $T$ as small as possible, we consider the computation of  $\Lambda_C$ without fully estimating $\mathbf d$. Further, since the coefficients  located far from the  first element of $\mathbf d$ can be negligible with  most of the  models considered  in \cite{Vuran_Else2004},  $\tilde{\boldsymbol\Sigma}_s$ reduces to a banded covariance matrix in which only  few off diagonals have significant coefficients. Thus, we expect that constructing $\Lambda_C$ estimating only  $1<K < L$ coefficients of $\mathbf d$ would  not result in  a significant performance degradation.

Let $\mathcal U_k$ be the set containing the all the $(i,j)$ pairs of the $k$-th diagonal in the upper triangle (including the main diagonal) of $\tilde{\boldsymbol\Sigma}_x$ for $k=0,1,\cdots,L-1$. It is noted that $\mathcal U_0$  corresponds to the main diagonal.
Let $\mathbf B_0=\underset{(i,i)\in \mathcal U_0}{\sum}{\mathbf a_i\mathbf a_i^T}$, $\mathbf B_k=\underset{(i,j)\in \mathcal U_k}{\sum}{\mathbf a_i\mathbf a_j^T + \mathbf a_j\mathbf a_i^T}$ for $k=1,\cdots, L-1$.  With the first $K$ significant elements of $\mathbf d$, $\tilde{\boldsymbol\Sigma}_y$ can be approximated by
\begin{eqnarray}
\tilde{\boldsymbol\Sigma}_y \approx \sum_{k=0}^{K-1} d_{k+1} \mathbf B_k.
\end{eqnarray}
While there are several approaches proposed in the literature to estimate the covariance matrix based on the compressed measurements \cite{Bioucas-Dias_EU2014,Wimalajeewa_Arx13,Romero_SPM16}, in this work, we consider the least squares (LS) method. Evaluation of the merits of different algorithms for covariance estimation is beyond the scope of this paper.
The  LS estimate of the first $K$ coefficients of $\mathbf d$, $\mathbf d_K$,  can be found as the solution to
\begin{eqnarray}
\hat{\mathbf d}_K = \underset{\mathbf d_K}{\arg\min}||\tilde{\boldsymbol\Sigma}_y  - \sum_{k=0}^{K-1} d_{k+1} \mathbf B_k  ||_F^2
\end{eqnarray}
which is given by,
\begin{eqnarray}
\hat{\mathbf d}_K = \mathbf H_K^{-1}\mathbf f_K
\end{eqnarray}
where $\mathbf H_K[i,j] = \mathrm{tr}(\mathbf B_{i-1}\mathbf B_{j-1}^T)$ for $i,j = 1\cdots, K$,  $\mathbf f_K[i] = \mathrm{tr}(\tilde{\boldsymbol\Sigma}_y \mathbf B_{i-1}^T)$ for $i = 1\cdots, K$,  $||\cdot||_F$ denotes the Frobebius norm and $\mathrm{tr}(\cdot)$ denotes the trace operator. Then, $\Lambda_C$ in \eqref{Lamda_C} can be approximated by,
\begin{eqnarray}
\Lambda_C \approx   \frac{L |\hat d_1| + 2\sum_{l=1}^{K-1}(L-l)|\hat d_{l+1}|}{L |\hat d_1| }.\label{Lamda_C_LS}
\end{eqnarray}

\section{Numerical Results}\label{sec_numerical}
To obtain  numerical results,  the random source is assumed to be Gaussian. We define the average SNR to be $\gamma_0 = 10\log_{10} \left(\frac{\sigma_S^2}{\sigma_{\tilde w}^2}\right)$.  We consider a scenario with $L$ equally spaced sensors in a 1-D space.
Further, we consider  the power exponential model for correlation \cite{Vuran_Else2004} in which $\rho_{k-1}$ in  $\mathbf d$ can be expressed as  $\rho_{k-1} = G_{\vartheta}(r_{1k})$ for $k=2,\cdots, L$ where $G_{\vartheta}(r_{1k}) = e^{-r_{1k}/\theta_1}$ for $\theta_1 > 0$. Let $r$ be the distance between any two sensors. Then, we can write $G_{\vartheta}(r_{1k}) = e^{-(k-1)r/\theta_1} = \left(e^{-r/\theta_1}\right)^{(k-1)} \triangleq \rho^{(k-1)}$ where $\rho=e^{-r/\theta_1}$ for $k=2.\cdots, L$. First, we select the elements of $\mathbf A$ so that $\mathbf A \mathbf A^T = \mathbf I$. With this selection, $\mathbf A$ is a dense matrix, thus, all the nodes transmit during each MAC transmission. The performance of the detector is evaluated via the probability of false alarm, $P_f$,  and probability of detection, $P_d$,  which are given by $
P_f = Pr(\Lambda_C \geq \tau_C | \mathcal H_0) $ and $P_d = Pr(\Lambda_C \geq \tau_C | \mathcal H_1)$,
respectively.

\begin{figure}[h!]
    \centering
     \begin{subfigure}[b]{0.38\textwidth}
        \includegraphics[width=\textwidth]{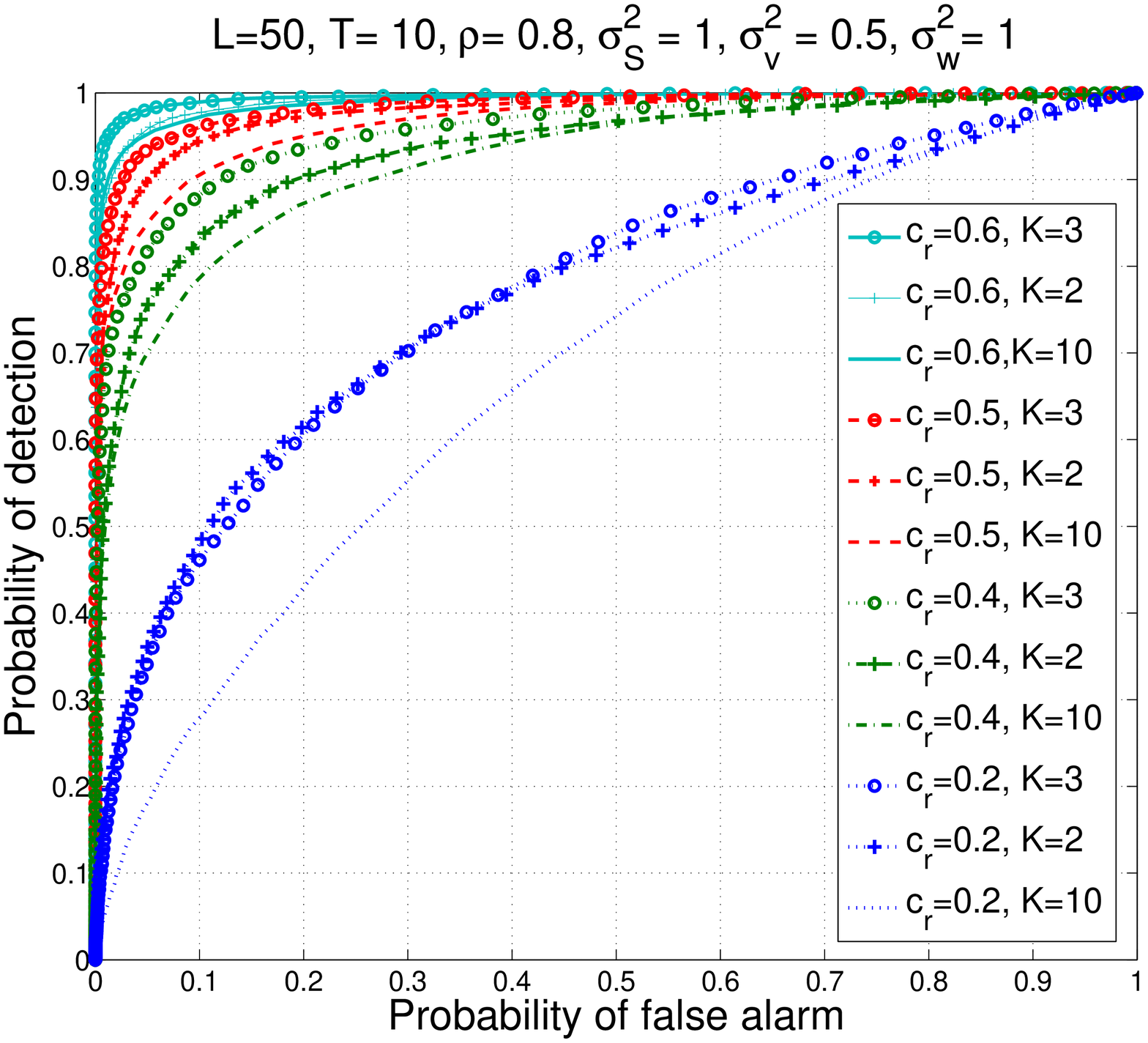}
        \caption{$T=10$}
        \label{fig:N1000}
    \end{subfigure}
    ~
    \begin{subfigure}[b]{0.38\textwidth}
        \includegraphics[width=\textwidth]{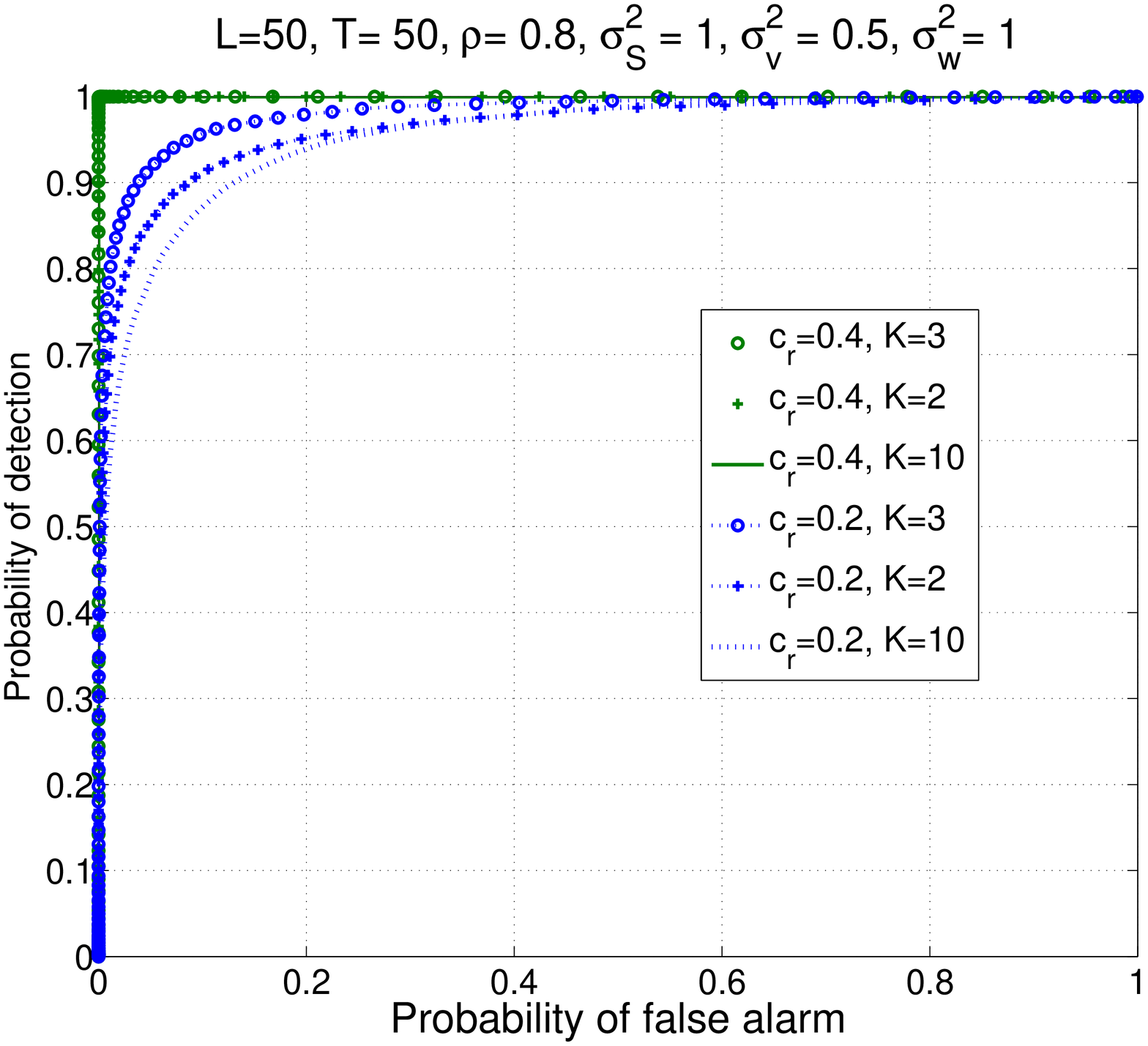}
        \caption{$T=50$}
        \label{fig:N100}
    \end{subfigure}
       \caption{Detection performance with $\Lambda_C$ as $c_r$ and $K$ vary for given $T$ and $L$, $L=50$ , $\rho=0.8$ }\label{fig_ROC_1}
\end{figure}

We show the detection performance with $\Lambda_C$ given in \eqref{Lamda_C_LS} in terms of ROC curves as $K$ varies for given $T$ and $L$ in Fig. \ref{fig_ROC_1}. We let $L=50$,  $\rho=0.8$, $\sigma_S^2=1$, $\sigma_v^2=0.5$, $\sigma_w^2=1$ so that  $\gamma_0 =  -1.7609 ~dB$. In Fig. \ref{fig_ROC_1}(a), $T=10$ while in Fig. \ref{fig_ROC_1} (b),  $T=50$. For given $T$ and $c_r \triangleq \frac{M}{L}$, it can be observed from Fig. \ref{fig_ROC_1}(a), and Fig. \ref{fig_ROC_1}(b) that, with large  $K$, the detection performance  degrades compared to relatively small $K$;  i.e., estimating only $K=3$ coefficients of $\mathbf d$ provides  better detection performance than that with $K=10$.  With limited $T$, when the number of elements to be estimated becomes larger, the error in estimation can increase, thus, performance with smaller $K$ is better than that with large $K$.  When $T$ increases from $T=10$ (Fig. \ref{fig_ROC_1} (a)) to $T=50$ (Fig. \ref{fig_ROC_1} (b)), improved performance for given $c_r$ is observed since then the sample estimate of $\tilde{\boldsymbol\Sigma}_y$ becomes more accurate  resulting in  a more accurate estimate for $\hat {\mathbf d}_K$.  In the  following figures, we set $K=3$ with $\Lambda_C$ unless otherwise specified.

\begin{figure}[h!]
\centerline{\epsfig{figure=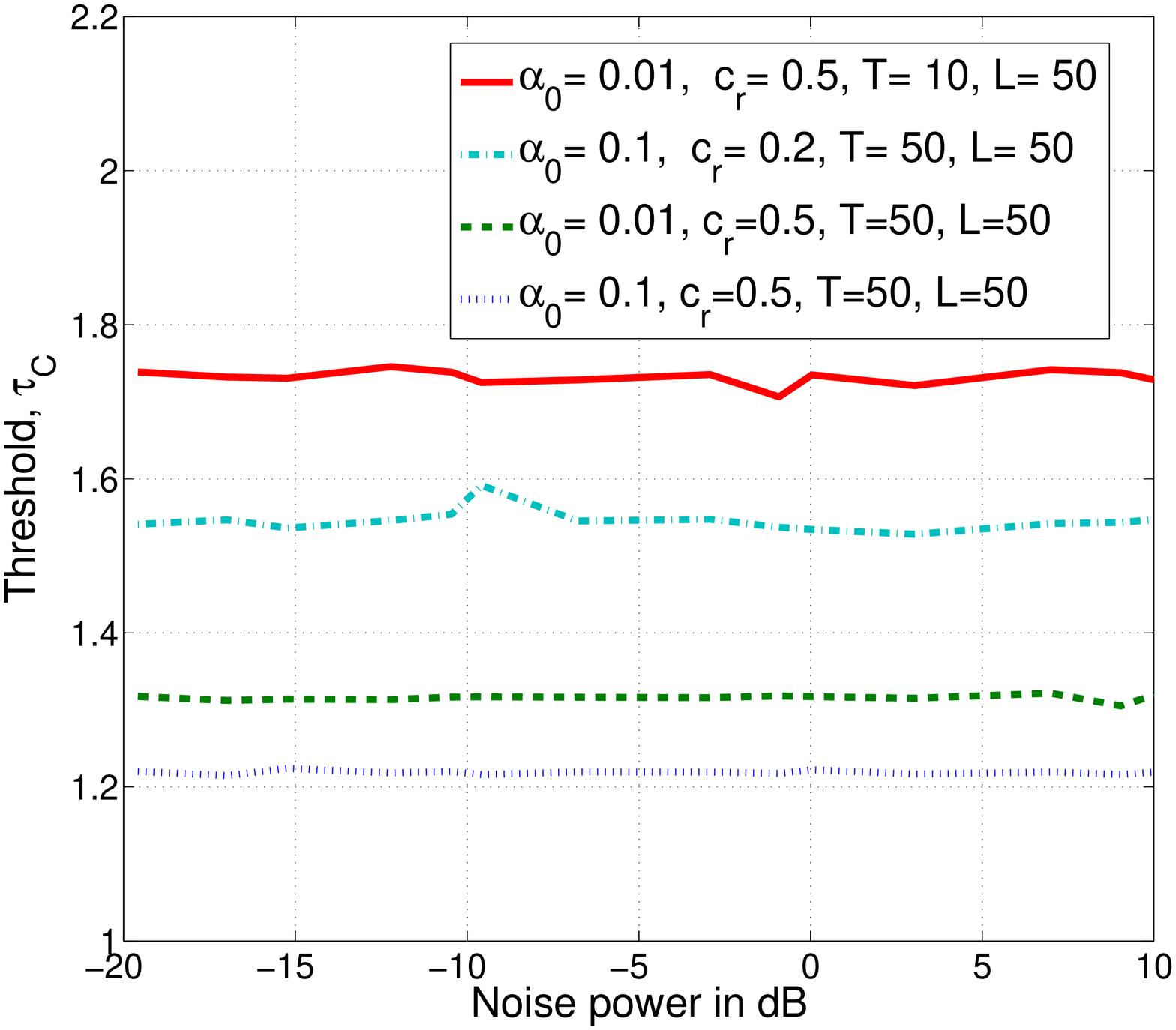,width=7.0cm}}
\caption{Threshold, to keep $P_f \leq \alpha_0$,  of the covariance based detector with $\Lambda_C$ vs $10\log_{10}{\sigma_{\tilde w}^2}$}\label{fig:threshold_cov}
\end{figure}
Let the desired probability of false alarm be  $\alpha_0$. In order to find the threshold of the detector with $\Lambda_C$, we need to find $\tau_C$ so that
$
Pr(\Lambda_C \geq \tau_C | \mathcal H_0) \leq \alpha_0
$,
which  is analytically difficult.  In Fig. \ref{fig:threshold_cov}, we plot $\tau_C$ computed numerically  as $\sigma_{\tilde w}^2$ varies keeping $L$, $T$ and $M$ fixed. The noise power along  the $x$-axis is taken as $ 10 \log_{10}(\sigma_{\tilde w}^2) $.  It can be observed that, the threshold is independent of the noise parameter for given $T$, $L$ and $c_r$ which makes the compressive covariance based detector attractive compared to the other non parametric detectors such as the energy detector.

\begin{figure}[h!]
    \centering
     \begin{subfigure}[b]{0.38\textwidth}
        \includegraphics[width=\textwidth]{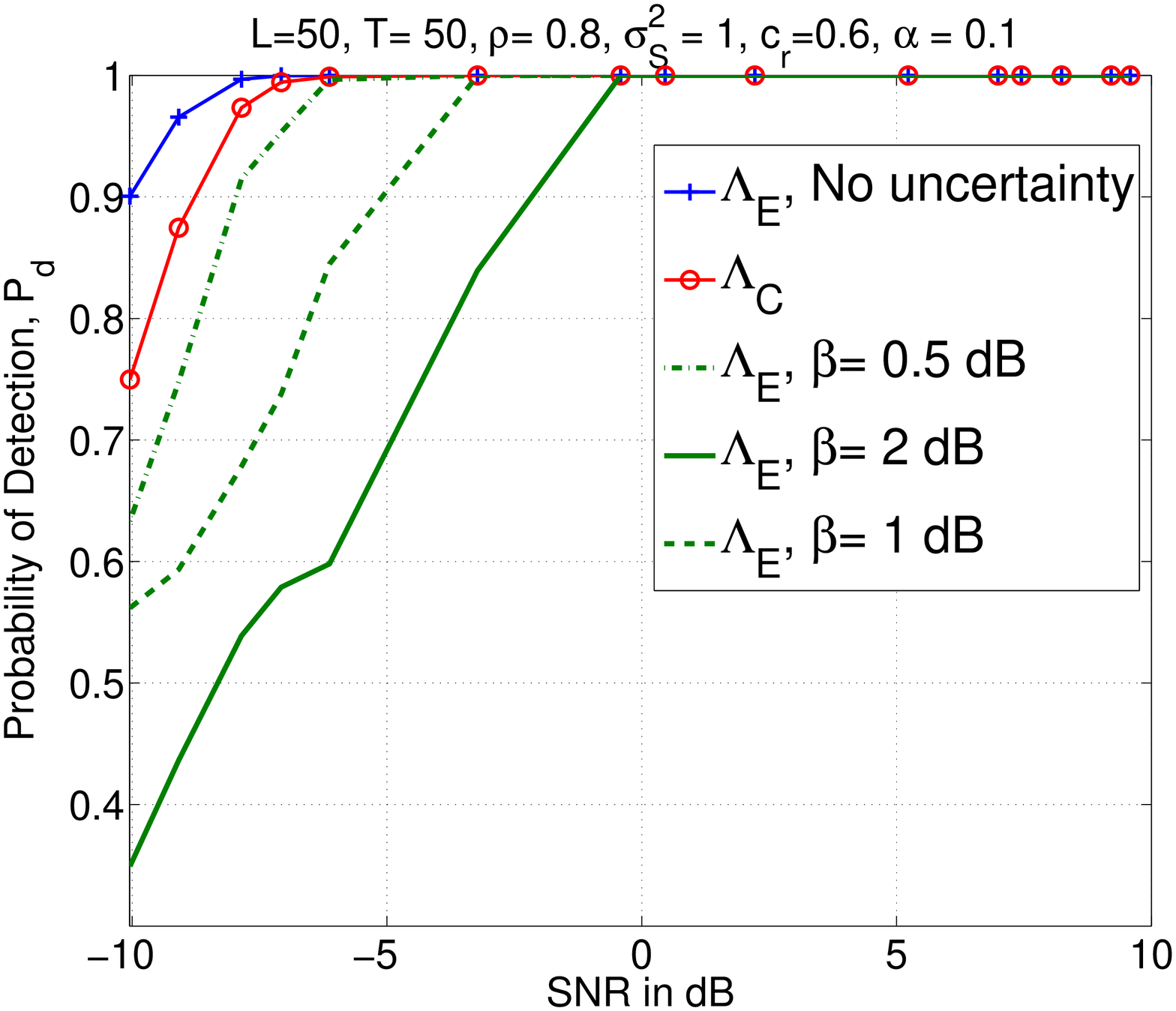}
        \caption{$P_d$}
        \label{fig:N1000}
    \end{subfigure}
    ~
    \begin{subfigure}[b]{0.38\textwidth}
        \includegraphics[width=\textwidth]{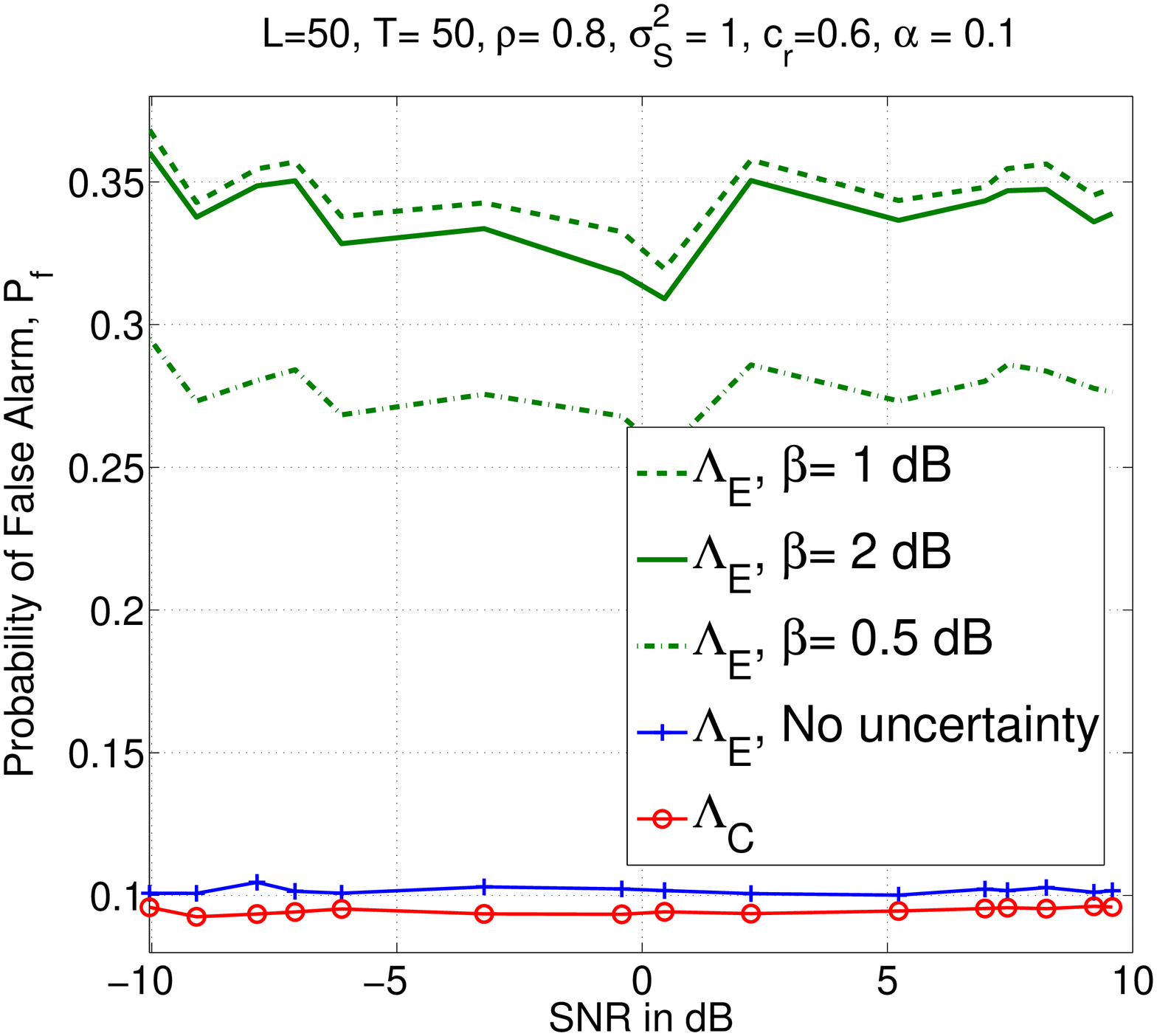}
        \caption{$P_f$}
        \label{fig:N100}
    \end{subfigure}
       \caption{Probability of detection and false alarm vs SNR }\label{fig_Pd_SNR}
\end{figure}

Next, we illustrate the robustness of the proposed detector compared to the energy detector. The  decision statistic of the  energy detector is given by
$
\Lambda_{E} = \sum_{n=1}^T ||\mathbf y[n]||^2
$. Approximating $\Lambda_{E}$ to be Gaussian under $\mathcal H_0$, the threshold of the energy detector to keep $P_f \leq \alpha_0$, $\tau_{E}$,  can be found as
$
\tau_E = \sigma_{\tilde w}^2 \left(\sqrt{2MT} Q^{-1} (\alpha_0) + MT\right)
$
which is a function of $\sigma_{\tilde w}^2$ where $Q^{-1}(\cdot)$ denotes the inverse Gaussian $Q$ function. The estimated or the assumed noise power in many practical receivers can be different from the real noise power.  Let $\tilde {\sigma}_{\tilde w}^2$ be the estimated noise power, which can be expressed as $\tilde{\sigma}_{\tilde w}^2=\beta_w \sigma_{\tilde w}^2$. The noise uncertainty factor is defined as $\beta=\max\{10\log_{10} \beta_w\}$ \cite{Zeng_C2007}. As in \cite{Zeng_C2007}, we assume that $\beta_w$ is uniformly distributed over  $[-\beta, \beta]$.
In Fig. \ref{fig_Pd_SNR},  $P_d$ and $P_f$ vs SNR are  plotted when detection is performed with $\Lambda_C$ and $\Lambda_E$ setting the threshold so the $\alpha_0=0.1$. To vary SNR, we vary $\sigma_v^2$ keeping $\sigma_S^2$ and $\sigma_w^2$ fixed.  With $\Lambda_C$, we compute the threshold numerically  for given $L$, $c_r$, and taking $\sigma_v^2=0.5$ and $\sigma_w^2=0.1$ and keep  it the same as SNR varies. With $\Lambda_E$, we plot $P_d$ and $P_f$ in the presence of noise variance  uncertainty (as $\beta$ varies)  as well as when it is assumed that there  in no uncertainty.

From Fig. \ref{fig_Pd_SNR}, it can be seen that when  there is no uncertainty in the estimated noise power, the energy detector has better detection performance than the covariance based detector. However, the  performance of the former, in terms of both $P_d$ and $P_f$,     degrades significantly even with small $\beta$.  Thus, detection based on $\Lambda_C$ appears to be more robust  in practical applications  than the energy detector.


\begin{figure}[h!]
\centerline{\epsfig{figure=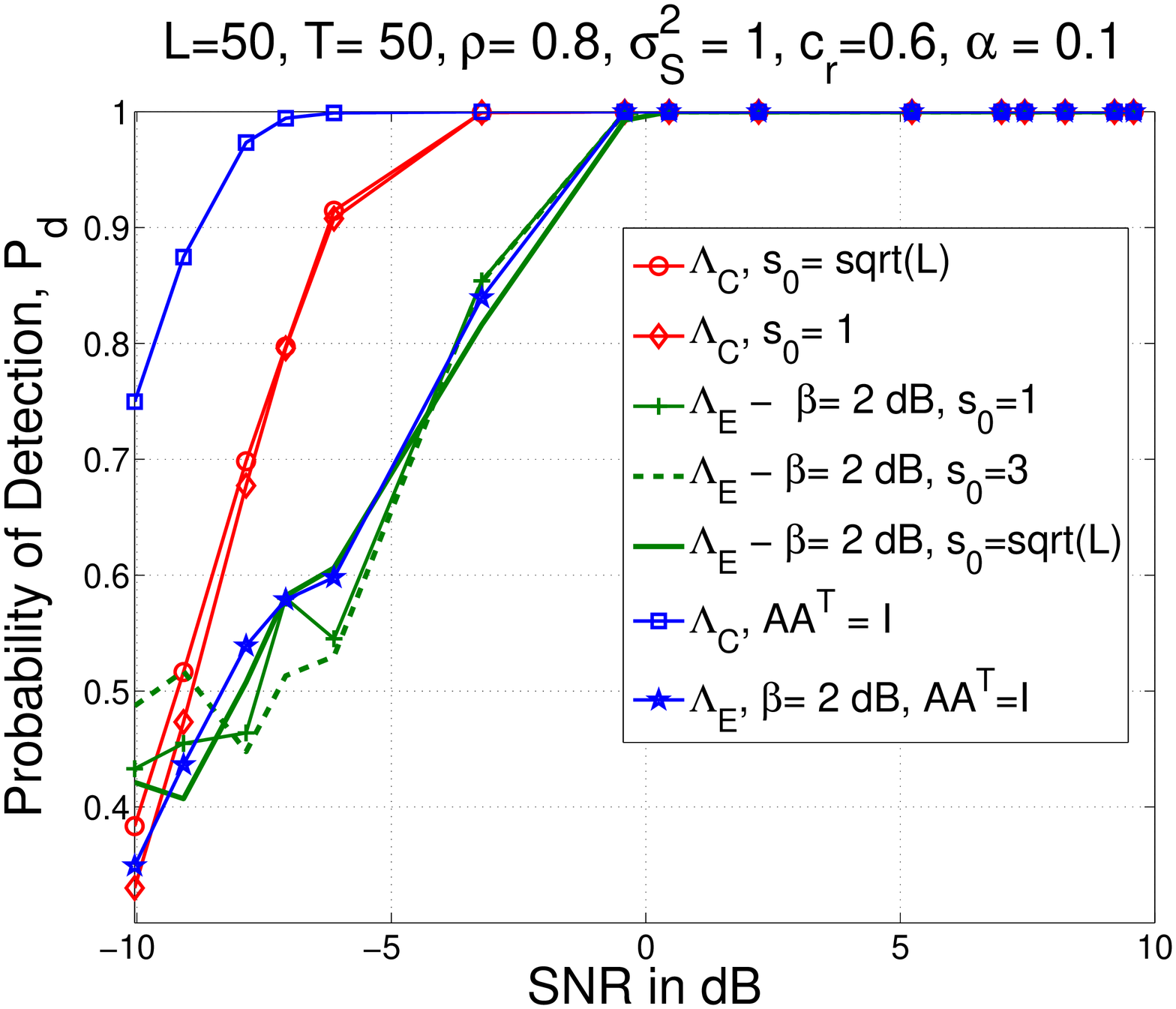,width=7.0cm}}
\caption{Probability of detection  vs SNR with sparse random projections}\label{fig_Pd_SNR_Sparse}
\end{figure}

Next, we investigate the detection performance when the assumption $\mathbf A \mathbf A^T = \mathbf I$ is relaxed. In resource constrained sensor networks, the use of sparse random projections for spatial data compression is promising \cite{wang_ISPN07,Yang_TSP2013,Wimalajeewa_TSIPN15} since then not all the sensors need to transmit during a given MAC transmission. To illustrate the detection performance, we select $\mathbf A[i,j]$ as
\begin{eqnarray}
\mathbf A[i,j] = \sqrt{\frac{s_0}{L}}\left\{
\begin{array}{cccc}
1 ~ & \mathrm{with} ~ \mathrm{prob} ~ \frac{1}{2 s_0}\\
0 ~ & \mathrm{with} ~ \mathrm{prob} ~ 1 - \frac{1}{s_0}\\
-1 ~ &  \mathrm{with} ~ \mathrm{prob} ~ \frac{1}{2 s_0}
\end{array}\right.\label{A_sparse}
\end{eqnarray}
with $s_0\geq 1$. With this matrix, only $L/s_0$ sensors, on an average, need to transmit during a given MAC transmission. When $s_0=1$, $\mathbf A$ is dense and all the nodes have to transmit.  In Fig. \ref{fig_Pd_SNR_Sparse}, we plot $P_d$ vs SNR as $s_0$ varies with $\Lambda_C$ and $\Lambda_E$ with $\beta= 2 dB$.  We let $\alpha_0=0.1$. We further plot the performance when $\mathbf A$ is selected such that $\mathbf A \mathbf A^T =\mathbf I$ as considered before so that $\tilde{\boldsymbol\Sigma}_x$ is exactly diagonal under $\mathcal H_0$. When comparing $\Lambda_C$ with $\mathbf A$ as in \eqref{A_sparse} for $s_0=1$, to $\Lambda_C$ with $\mathbf A \mathbf A^T =\mathbf I$, it can be seen from Fig. \ref{fig_Pd_SNR_Sparse} that the former provides with  a degraded performance compared  to the latter. This is due to the fact that, with the former, $\tilde{\boldsymbol\Sigma}_x$ is only approximately diagonal under $\mathcal H_0$ which reduces the  distinguishability between the two hypotheses. However, compared to the energy detector with noise uncertainty,  $\Lambda_C$ with $\mathbf A$ as in \eqref{A_sparse} even with very small $1/s_0$ provides much better detection performance. Further, it is seen that the sparsity parameter of $\mathbf A$ in \eqref{A_sparse}, $s_0$,  does not impact on the detection performance significantly. Thus, it is sufficient for only a small number of nodes (e.g., $\sqrt {L}$ on average) to transmit observations to achieve almost the same performance as when all the $L$ nodes  transmit with the matrix $\mathbf A$ in \eqref{A_sparse}.


\section{Conclusion}\label{sec_conclusion}
In this paper, we have proposed a nonparametric detection method exploiting CS to detect a random event using spatially correlated data in a sensor network. To transmit  a compressed version of spatially correlated data at the fusion center, the MAC model was employed. A test statistic based on the covariance  matrix of uncompressed data was  considered which was computed based on the limited compressed samples received at the fusion center.  Unlike the widely used energy detector, the proposed detector does not need exact estimates of the noise power to set the threshold. Further, the proposed detector is robust to  the selection of the sparsity parameter of the random projection matrix when sparse random projections are employed to reduce the communication overhead.

\newpage
\newpage

\bibliographystyle{IEEEtran}
\bibliography{IEEEabrv,bib1,ref_1,ref_SurveyP16}

\begin{thebibliography}{10}
\providecommand{\url}[1]{#1}
\csname url@samestyle\endcsname
\providecommand{\newblock}{\relax}
\providecommand{\bibinfo}[2]{#2}
\providecommand{\BIBentrySTDinterwordspacing}{\spaceskip=0pt\relax}
\providecommand{\BIBentryALTinterwordstretchfactor}{4}
\providecommand{\BIBentryALTinterwordspacing}{\spaceskip=\fontdimen2\font plus
\BIBentryALTinterwordstretchfactor\fontdimen3\font minus
  \fontdimen4\font\relax}
\providecommand{\BIBforeignlanguage}[2]{{%
\expandafter\ifx\csname l@#1\endcsname\relax
\typeout{** WARNING: IEEEtran.bst: No hyphenation pattern has been}%
\typeout{** loaded for the language `#1'. Using the pattern for}%
\typeout{** the default language instead.}%
\else
\language=\csname l@#1\endcsname
\fi
#2}}
\providecommand{\BIBdecl}{\relax}
\BIBdecl

\bibitem{Akyildiz_CN2002}
I.~Akyildiz, W.~Su, Y.~Sankarasubramaniam, and E.~Cayirci, ``Wireless sensor
  networks: a survey,'' \emph{Computer Networks}, vol.~38, no.~4, 2002.

\bibitem{Puccinelli_MCAS2005}
D.~Puccinelli and M.~Haenggi, ``Wireless sensor networks: Applications and
  challenges of ubiquitous sensing,'' \emph{{IEEE} Circuits Syst. Mag.},
  vol.~5, no.~3, pp. 19--31, 2005.

\bibitem{Yick_CN2008}
J.~Yick, B.~Mukherjee, and D.~Ghosal, ``Wireless sensor network survey,''
  \emph{Computer Networks}, vol.~52, no.~12, pp. 2292--2330, 2008.

\bibitem{Mainetti_SoftCOM2011}
L.~Mainetti, L.~Patrono, and A.~Vilei, ``Evolution of wireless sensor networks
  towards the internet of things: A survey,'' in \emph{19th International
  Conference on Software, Telecommunications and Computer Networks (SoftCOM)},
  2011, pp. 1--6.

\bibitem{Stankovic_CSN2011}
J.~A. Stankovic, A.~D. Wood, and T.~He, ``Realistic applications for wireless
  sensor networks,'' \emph{Theoretical Aspects of Distributed Computing in
  Sensor Networks Springer}, pp. 853--863, 2011.

\bibitem{Othman_EP2012}
M.~F. Othman and K.~Shazali, ``Wireless sensor network applications: A study in
  environment monitoring system,'' \emph{Engineering Procedia}, vol.~41, pp.
  1204--1210, 2012.

\bibitem{Rawat_JoS2014}
P.~Rawat, K.~D. Singh, H.~Chaouchi, and J.~M. Bonnin, ``Wireless sensor
  networks: a survey on recent developments and potential synergies,''
  \emph{Journal of supercomputing}, vol.~68, no.~1, pp. 1--48, 2014.

\bibitem{Rashid_JNCA2016}
B.~Rashid and M.~H. Rehmani, ``Applications of wireless sensor networks for
  urban areas: A survey,'' \emph{Journal of Network and Computer Applications},
  vol.~60, pp. 192--219, 2016.

\bibitem{Duarte_TIT13}
M.~F. Duarte, M.~B. Wakin, D.~Baron, S.~Sarvotham, and R.~G. Baraniuk,
  ``Measurement bounds for sparse signal ensembles via graphical models,''
  \emph{{IEEE} Trans. Inf. Theory}, vol.~59, no.~7, pp. 4280--4289, Jul. 2013.

\bibitem{haupt_SPM2008}
J.~Haupt, W.~U. Bajwa, M.~Rabbat, and R.~Nowak, ``Compressed sensing for
  networked data,'' \emph{{IEEE} Signal Process. Mag.}, vol.~25, no.~2, pp.
  92--101, Mar. 2008.

\bibitem{meng_CISS09}
J.~Meng, H.~Li, and Z.~Han, ``Sparse event detection in wireless sensor
  networks using compressive sensing,'' in \emph{43rd Annual Conf. on
  Information Sciences and Systems (CISS)}, Baltimore, MD, Mar. 2009, pp. 181
  -- 185.

\bibitem{Feng_Globecom2009}
C.~Feng, S.~Valaee, and Z.~H. Tan, ``Multiple target localization using
  compressive sensing,'' in \emph{IEEE Global Telecommunications Conference
  (GLOBECOM)}, Dec. 2009.

\bibitem{Zhang_Infocom2011}
B.~Zhang, X.~Cheng, N.~Zhang, Y.~Cui, Y.~Li, and Q.~Liang, ``Sparse target
  counting and localization in sensor networks based on compressive sensing,''
  in \emph{INFOCOM}, 2011.

\bibitem{duarte_ICASSP06}
M.~F. Duarte, M.~A. Davenport, M.~B. Wakin, and R.~G. Baraniuk, ``Sparse signal
  detection from incoherent projections,'' in \emph{Proc. Acoust., Speech,
  Signal Processing (ICASSP)}, May 2006.

\bibitem{haupt_ICASSP07}
J.~Haupt and R.~Nowak, ``Compressive sampling for signal detection,'' in
  \emph{Proc. Acoust., Speech, Signal Processing (ICASSP)}, vol.~3, Honolulu,
  Hawaii, Apr. 2007, pp. III--1509 -- III--1512.

\bibitem{Gang_globalsip14}
G.~Li, H.~Zhang, T.~Wimalajeewa, and P.~K. Varshney, ``On the detection of
  sparse signals with sensor networks based on \textsc{S}ubspace
  \textsc{P}ursuit,'' in \emph{IEEE Global Conference on Signal and Information
  Processing (GlobalSIP)}, Atlanta, GA, Dec. 2014, pp. 438--442.

\bibitem{Rao_icassp2012}
B.~S. M.~R. Rao, S.~Chatterjee, and B.~Ottersten, ``Detection of sparse random
  signals using compressive measurements,'' in \emph{Proc. Acoust., Speech,
  Signal Processing (ICASSP)}, 2012, pp. 3257--3260.

\bibitem{Cao_Info2014}
J.~Cao and Z.~Lin, ``Bayesian signal detection with compressed measurements,''
  \emph{Information Sciences}, pp. 241--253, 2014.

\bibitem{Wimalajeewa_tsipn16}
T.~Wimalajeewa and P.~K. Varshney, ``Sparse signal detection with compressive
  measurements via partial support set estimation,'' \emph{IEEE Trans. on
  Signal and Inf. Process. over Netw.}, vol.~3, no.~1, Mar. 2017.

\bibitem{davenport_JSTSP10}
M.~A. Davenport, P.~T. Boufounos, M.~B. Wakin, and R.~Baraniuk, ``Signal
  processing with compressive measurements,'' \emph{IEEE J. Sel. Topics Signal
  Process.}, vol.~4, no.~2, pp. 445 -- 460, Apr. 2010.

\bibitem{Wimalajeewa_asilomar10}
T.~Wimalajeewa, H.~Chen, and P.~K. Varshney, ``Performance analysis of
  stochastic signal detection with compressive measurements,'' in
  \emph{$44^{\mathrm{th}}$ Annual Asilomar Conf. on Signals, Systems and
  Computers}, Nov. 2010, pp. 913--817.

\bibitem{Bhavya_cscps14}
B.~Kailkhura, T.~Wimalajeewa, L.~Shen, and P.~K. Varshney, ``Distributed
  compressive detection with perfect secrecy,'' in \emph{2nd Int. Workshop on
  Compressive Sensing in Cyber-Physical Systems (CSCPS'14)}, Oct. 2014.

\bibitem{Bhavya_asilomar14}
B.~Kailkhura, T.~Wimalajeewa, and P.~K. Varshney, ``On physical layer secrecy
  of collaborative compressive detection,'' in \emph{$48^{\mathrm{th}}$ Annual
  Asilomar Conf. on Signals, Systems and Computers}, 2014.

\bibitem{Vuran_Else2004}
M.~Vuran, O.~Akan, and I.~Akyildiz, ``Spatio-temporal correlation: Theory and
  applications for wireless sensor networks,'' \emph{Comput. Networks
  (Elsevier)}, vol.~45, no.~3, pp. 245--259, 2004.

\bibitem{Berger_Stat2001}
J.~Berger, V.~de~Oliviera, and B.~Sanso, ``Objective bayesian analysis of
  spatially correlated data,'' \emph{J. Am. Statist. Assoc.}, vol.~96, pp.
  1361--1374, 2001.

\bibitem{Romero_SPM16}
D.~Romero, D.~Ariananda, Z.~Tian, and G.~Leus, ``Compressive covariance
  sensing: Structure-based compressive sensing beyond sparsity,'' \emph{{IEEE}
  Signal Process. Mag.}, vol.~33, no.~1, pp. 78--93, Jan. 2016.

\bibitem{Bajwa_IT2007}
W.~Bajwa, J.~Haupt, A.~Sayeed, and R.~Nowak, ``Joint source–channel
  communication for distributed estimation in sensor networks,'' \emph{{IEEE}
  Trans. Inf. Theory}, vol.~53, no.~10, pp. 3629--3653, Oct. 2007.

\bibitem{Wimalajeewa_TSP17}
T.~Wimalajeewa and P.~K. Varshney, ``Compressive sensing based detection with
  multimodal-dependent data,'' \emph{Online
  available,https://arxiv.org/pdf/1701.01352.pdf}, 2017.

\bibitem{Zeng_C2007}
Y.~Zeng and Y.-C. Liang, ``Covariance based signal detections for cognitive
  radio,'' in \emph{IEEE Int. Symposium on New Frontiers in Dynamic Spectrum
  Access Networks}, Dublin, Ireland, Apr. 2007, pp. 202--207.

\bibitem{Zeng_VT09}
------, ``Spectrum-sensing algorithms for cognitive radio based on statistical
  covariances,'' \emph{{IEEE} Trans. Veh. Technol.}, vol.~58, no.~4, pp.
  1804--1815, May 2009.

\bibitem{Bioucas-Dias_EU2014}
J.~Bioucas-Dias, D.~Cohen, , and Y.~Eldar, ``Covalsa: Covariance estimation
  from compressive measurements using alternating minimization,'' in \emph{2014
  Proceedings of the 22nd European Signal Processing Conference (EUSIPCO)},
  Sept. 2014, pp. 999--1003.

\bibitem{Wimalajeewa_Arx13}
T.~Wimalajeewa, Y.~C. Eldar, and P.~K. Varshney, ``Recovery of sparse matrices
  via matrix sketching,'' \emph{Co\textsc{RR}, vol. abs/1311.2448}, 2013.

\bibitem{wang_ISPN07}
W.~Wang, M.~Garofalakis, and K.~Ramchandran, ``Distributed sparse random
  projections for refinable approximation,'' in \emph{ISPN}, Cambridge,
  Massachusetts,USA, April 2007, pp. 331--339.

\bibitem{Yang_TSP2013}
G.~Yang, V.~Tan, C.~Ho, S.~Ting, and Y.~Guan, ``Wireless compressive sensing
  for energy harvesting sensor nodes,'' \emph{IEEE Trans. Signal Process.},
  vol.~61, p. 4491 – 4505, Sept. 2013.

\bibitem{Wimalajeewa_TSIPN15}
T.~Wimalajeewa and P.~K. Varshney, ``Wireless compressive sensing over fading
  channels with distributed sparse random projections,'' \emph{IEEE Trans.
  Signal and Inf. Process. over Netw}, vol.~1, no.~1, Mar. 2015.

\end{thebibliography}

\end{document}